# From Questions to Effective Answers: On the Utility of Knowledge-Driven Querying Systems for Life Sciences Data

Amir H. Asiaee[1], Prashant Doshi[1], Todd Minning[2], Satya Sahoo[3], Priti Parikh[3], Amit Sheth[3] and Rick L. Tarleton[2]

[1] THINC Lab, Dept. of Computer Science, University of Georgia, Athens, GA
[2] Tarleton Research Group, Dept. of Cellular Biology, University of Georgia, Athens, GA
[3] Kno.e.sis Center, Dept. of Computer Science, Wright State University, Dayton, OH
aha@uga.edu, pdoshi@cs.uga.edu, {tminning,tarleton}@uga.edu,
{satya,priti,amit}@knoesis.org

**Abstract.** We compare two distinct approaches for querying data in the context of the life sciences. The first approach utilizes conventional databases to store the data and intuitive form-based interfaces to facilitate easy querying of the data. These interfaces could be seen as implementing a set of "pre-canned" queries commonly used by the life science researchers that we study. The second approach is based on semantic Web technologies and is knowledge (model) driven. It utilizes a large OWL ontology and same datasets as before but associated as RDF instances of the ontology concepts. An intuitive interface is provided that allows the formulation of RDF triples-based queries. Both these approaches are being used in parallel by a team of cell biologists in their daily research activities, with the objective of gradually replacing the conventional approach with the knowledge-driven one. This provides us with a valuable opportunity to compare and qualitatively evaluate the two approaches. We describe several benefits of the knowledge-driven approach in comparison to the traditional way of accessing data, and highlight a few limitations as well. We believe that our analysis not only explicitly highlights the specific benefits and limitations of semantic Web technologies in our context but also contributes toward effective ways of translating a question in a researcher's mind into precise computational queries with the intent of obtaining effective answers from the data. While researchers often assume the benefits of semantic Web technologies, we explicitly illustrate these in practice.

## 1 Introduction

The life sciences – almost uniquely – generate and manage very large amounts of data using complex processes. These data range from being genomic and proteomic to procedural and in contexts ranging from parasites to humans. It is not surprising, then, that semantic Web technologies are finding pervasive applications in the life sciences. Predominant, among the many, is the use of standard ontologies (e.g., BioPortal at the National Center for Biomedical Ontologies [1]) to structure biomedical knowledge, and semantic data models such as RDF [2]. Both these semantic Web technologies allow life science researchers to manage and exchange data in collectively understandable formats. This is a significant benefit that facilitates scientific collaborations and dissemination, potentially leading to quicker progress.

Much of the data in the life sciences continues to be stored using conventional database management systems (DBMS) and subsequently, queried using the structured query language (SQL) supported by these DBMSs. Intuitive interfaces such as forms either available on the Web or on local intranets facilitate entering and querying the data. Often, these interfaces support "pre-canned" queries that are most commonly used by the researchers. The efficiency of modern DBMSs and the intuitive nature of the interfaces together make this approach adequate for the researchers who are chiefly interested in quick and targeted accessibility to the data. However, the static interfaces often tend to throw up more data than needed leading to time-consuming post processing steps, and the tabular schemas do not make the conceptual relationships explicit making queries specific to the local setup and researchers, instead of being general.

We compare and contrast two approaches for querying life sciences data. Both these approaches utilize an identical data context: strain, stage transcriptome and proteomic data on the parasite Trypanosoma cruzi (*T.cruzi*). This parasite is responsible for the *Chagas* disease that is prevalent throughout Latin America and is often fatal. In the first approach, *T.cruzi* data is stored in a conventional DBMS and accessed through a suite of well-designed forms, which essentially represent a predefined set of commonly used queries. We refer to this approach as **Paige Tools** [3] after the name of the server that hosts these forms. **Paige Tools** has been the de-facto way for storing and accessing experimental data related to *T.cruzi* by the Tarleton research group located in the Center for Tropical and Emerging Diseases at the University of Georgia.  The second approach uses an OWL-based ontology designed in collaboration with the life science researchers in order to model the experimental data related to *T.cruzi*. The ontology supports the data modeled using RDF. Querying capabilities are provided by a significantly enhanced version of the knowledge-driven querying system, **Cuebee** [4][15]. It provides an intuitive interface that facilitates formulation of RDF triples-based queries, which are then transformed into SPARQL-DL [6]. Previously, Mendes et al. [15] introduced **Cuebee** [4] and demonstrated its preliminary use in the context of *T.cruzi*. In this paper, we explicitly illustrate four benefits (and two limitations) of enhanced **Cuebee** related to its usefulness that arise on deployment, using concrete examples.

We think that **Paige Tools** and **Cuebee** are representative of the traditional and more sophisticated way of querying life sciences data, respectively. These approaches provide alternative ways of transforming the precise question in a researcher's mind into a computational query and then obtaining the solution to the query, which forms the answer. The outcome of our analysis is a set of benefits that knowledge-driven approaches such as Cuebee offer over the more conventional approaches. We also highlight two limitations that this approach faces, which could impede its widespread adoption despite the substantial benefits.

The rest of the paper is structured as follows: In Section 2, we discuss other systems utilizing semantic Web technologies in the life science and other contexts. Section 3 describes Paige Tools and Cuebee approaches in more detail. In Section 4, we demonstrate the benefits of using the knowledge-driven query approach, Cuebee, over the conventional approach. Section 5 balances this by emphasizing a few limitations of approaches such as Cuebee in our context of life sciences. We conclude this paper with a discussion of our evaluation and its implications in Section 6.

## 2    Related Work

Other semantic Web based systems exist that focus on queries to provide targeted access to data in the life sciences and other contexts. These include query tools such as Openlink iSPARQL [7] and NITELIGHT [8] both of which provide graph-based interfaces for query formulation. A user generates a visual graph by adding concepts and connecting them together using relationships. iSPARQL is freely available and Kiefer et al. [33] evaluate it on a single data set. However, to the best of our knowledge, none of these systems evaluated their usefulness on use cases or are in use. Similar to Cuebee, GINSENG [9] offers suggestions to users but from a different perspective. GINSENG relies on a simple question grammar, which is extended using the ontology schema to guide users to directly formulate SPARQL queries. Bernstein et al. [9] briefly evaluated GINSENG on three aspects: usability of the system in a realistic task, its ability to parse large numbers of real-world queries, and its query performance. The experimental results did not compare GINSENG to other systems, and no real-world use of the system has been reported.

Semantics-based approaches also exist that focus more on the data integration aspect rather than query in the life sciences context. GoWeb [10] is a semantic search engine for the life sciences which combines classical keyword-based Web search with text-mining and ontologies to explore result sets and facilitate question answering. Dietze et al. [10] evaluated GoWeb on three benchmarks: BioCreAtIvE 1 (Task 2) [30] in the context of genes and functions, the study by Tang et al. [31] in the context of symptoms and diseases, and the questions from the 2006 TREC Genomics Track [32]. GoWeb provided answers with a recall of 58.1%, 77%, and 78.6% respectively. BioGateway [12] composes several online (such as OBO foundry [13] and GO annotation files [14]) and in house data sources, and provides a single entry point to query through SPARQL. Cheung et al. [11] introduce semantic Web query federation in the context of neuroscience. Their approach focuses on providing facilities to integrate different data sources and offers either SPARQL or SQL query interfaces to access remote data. However, the usefulness of the system has not been demonstrated in a real-world context. While the above systems operating in the context of life science data are available for public use, we did not find evidence of these systems being used by life science researchers. Furthermore, there is a general lack of explicit comparisons between these approaches and traditional systems. Thus, while Cuebee is not alone in its effort to bring knowledge-driven approaches to the life sciences, we believe that our comparative case study of the system in use is novel.

Mendes et al. [4][15] introduced Cuebee and demonstrated it in the context of *T.cruzi* data. They showed that its *usability* was comparable to existing DBMS based systems. This paper briefly discusses the enhancements to Cuebee, and explicitly illustrates the benefits and limitations of the *usefulness* of Cuebee while being used by an interdisciplinary team of computer science and cell biology researchers.

## 3   Background

In this section we briefly describe the two approaches for storing and querying experimental data related to *T.cruzi*. We emphasize that both Paige Tools and Cuebee are currently operational and are being used by researchers, with the expected longer-term objective of replacing Paige Tools with Cuebee.

### 3.1   Paige Tools – Conventional DBMS-based Approach

Paige Tools offers interfaces to add and edit experimental data related to *T.cruzi* housed in multiple separate local databases as well as facilities to execute queries over the stored data. These interfaces are available on the Web and are served through the lab website to the researchers. Our focus is on a subset of interfaces, which allow storing and querying of data in the context of the gene knockout protocol and parasite strains protocol, accompanied and annotated by experimental data. Access to the data is spread across three forms requiring the user to select one of them based on which dataset she intends to query.

Typically, these interfaces manifest as forms containing popular widgets such as drop-down lists, check boxes and buttons. While the drop-down lists and buttons allow the formulation of a boolean query on a specific dataset for each interface, the check boxes allow the selection of associated attributes to display in the result. Input from each form is transformed into a SQL query. However, researchers using these forms need not have any knowledge of SQL, and this is often the case. Output of SQL query is shown to user in a tabular format.

We believe that the interfaces in Paige Tools are typical of systems utilized by life science researchers. As expressed by the researchers that use Paige Tools, these tend to be simple but adequate approaches for somewhat targeted access to portions of data. The interfaces are tightly coupled to the schema design and users are limited to executing a set of queries allowed by the interface. One of these is usually a query that throws up all the data in the dataset. Due to the tight coupling, any change to the database schema results in refactoring of the forms to support the changes.

### 3.2  Cuebee – Knowledge-Driven Approach

Cuebee is an ontology-based query formulation and data retrieval system applied in the context of *T.cruzi* research. We enhanced the original system as described by Mendes et al. [4][15] with significant infrastructural modifications and new functionality. We begin by describing the original system followed by our enhancements in the next two subsections.

**2.2.1 Preliminary System.**   Cuebee allows querying of data modeled using RDF. This data could be housed in conventional DBMSs but published in RDF using D2R [16] or directly available in the RDF model. The original version of Cuebee [4] in the context of *T.cruzi* [15], utilized the latter setup. Because the RDF data is accessed using query endpoints, new data sources may be added in a plug-and-play manner without much developmental effort.

Query formulation within Cuebee utilizes ontology schemas to guide a user through the process of transforming her question into a query in a logical way. These queries are formulated as RDF triples (subject → relation → object), which could be arbitrarily long. Internally, the triples are transformed into SPARQL [17] queries which are executed using the Joseki server [18].

What sets Cuebee apart from other RDF-based querying tools is its *suggestion engine*[1]. It utilizes ontology schemas designed in RDFS to suggest concepts in a drop-down list that match the characters that the user starts typing. Furthermore, it lists all the relationships that are relevant for any particular concept selected by the user. Both these features reduce the need for users to be *a'priori* acquainted with the ontology – a major concern for ontology-driven systems. This process of formulating concepts and relationships represents an intuitive way of formulating an expressive computational query from the original question in the researcher's mind.

The *suggestion engine* supports these features by rapidly querying only the ontology schema and displaying the results. Consequently, each dataset should be accompanied by a schema, and the ontology schemas should be setup as distinct SPARQL endpoints; therefore, for each data source two SPARQL endpoints are employed: one for the data and the other for the schema. The *suggestion engine* uses the ontology schema endpoint only.

Both the query engines are implemented on the client side of a Web-based interface with which users interact. Communication between the client-side query engines and the server-side SPARQL endpoints where the ontology schemas and datasets reside is established through the SPARQL protocol for RDF [19]. In order to execute each query generated by either the suggestion or the answer engine, the client-side Web interface sends AJAX asynchronous calls. Then, SPARQL endpoints send back the results to the interface. These results are parsed and displayed to the users using different visualization methods such as tables, pie-charts or graphs [15].

---

[1] We introduce this nomenclature for notational convenience – it is not used by Mendez et al. [15]

**3.2.2    Revisions and Enhancements.**   We introduced several infrastructural modifications and enhancements to the preliminary version of Cuebee described previously. These include additional support for OWL-based ontologies, interface enhancements to support improved query formulation and display experience, and integration of Web services to enrich some of the final results with operations on external data sources. The enhanced version is available for use at [5].

The enhanced Cuebee allows the same steps of guiding users through ontology schemas in order to generate queries as the preliminary system.  However, we introduce multiple enhancements to the Web-based interface to make it more user-friendly. For example, Cuebee now annotates each suggested concept with information that includes a description of the concept, alternate labels if any and associated properties, in a pop-up. It allows selection of multiple instances that satisfy Boolean operators. All of this information is obtained from the ontology by the *suggestion engine*. In addition, an *undo* feature helps users revise their queries at any point during the query formulation process and after answers have been generated.

Additionally, our contributions go beyond the interface and focus on the infrastructure of Cuebee as well. A major improvement to the preliminary system is the addition of the capability to support OWL ontologies because OWL-based ontologies tend to be more expressive than those in plain RDFS, in part due to the use of restrictions.  For example, in the context of *T.cruzi* research, we use the OWL-based parasite experiment and life cycle ontologies [23] in Cuebee. Subsequently, we equip the two query engines to execute SPARQL-DL [6] queries which offer more expressive querying than SPARQL. In particular, they allow integration of ABox and TBox queries in a single SPARQL-DL query. OWL ontologies are deployed in a popular OWL-DL reasoner called Pellet [20] in order to take advantage of the inferencing capabilities offered by a powerful ontology reasoner. A secondary benefit of Pellet is that we can consolidate the two SPARQL endpoints, one for ontology schema and the other for the associated dataset, as required in the preliminary system into a single endpoint. As a result, querying multiple data sources becomes more straightforward. Finally, we significantly revised the automatic generation of the SPARQL-DL queries to include concepts and properties defined using restrictions (b-nodes) in the OWL ontologies.

In life sciences there are a large number of bioinformatics tools and data sources available as Web services (for e.g., see BioCatalogue [21]).  These often give access to large community data sets and are indispensible to the life science researcher. One such Web service is the NCBI BLAST [22] which allows the retrieval of aligned sequences by searching over large datasets using BLAST – an algorithm for comparing primary biological sequence information, such as the amino-acid sequences of different proteins or the nucleotides of DNA sequences. As another contribution to the preliminary system, we extend the results of the final queries with bioinformatics tools such as NCBI BLAST available as RESTful Web services. In this case, we detect if the results of any query contain appropriate types of sequences, and allow the user to trigger an asynchronous invocation of the NCBI BLAST Web service. This retrieves the BLAST results remotely, which are then displayed to the user in an informative manner.

## 4   Benefits of Cuebee over Paige Tools

Both Paige Tools and the enhanced Cuebee are running concurrently since the last six months; they provide access to identical data and both are in use by a team of cell biology researchers. The identical contexts provide us with a valuable opportunity to comparatively evaluate the two approaches in a principled way. We think that approaches such as Cuebee provide *four* significant benefits over traditional approaches, which we describe in this section. However, its usefulness also suffers from *two* limitations outlined in the next section.

### 4.1   Explicitly Structured Queries

The first benefit is with respect to the structure of the queries that may be formulated in the two approaches. In order to illustrate this, consider the following question posed by our team of life science researchers in the context of *T.cruzi*:

*Which microarray oligonucleotides from homologous genes have 3 prime region primers*?

In the above question, note that homology is a relationship between two genes (these genes are derived from a common ancestor) and 3-prime-region is a property of primers.

Conventional database design places minimal importance on named relationships between concepts (for e.g., table joins), and the underlying database structure in Paige Tools reflects this. While query pages within Paige Tools provide users the ability to show attributes of microarray oligonucleotide, genes and primers, discerning any homology relationships between two gene sequences or whether a primer has a 3 prime region is left to the ability of the user. In their use of Paige Tools, researchers imply these relationships using a series of post-processing steps on the results. Thus, the resulting query does not adequately reflect the original question in the researcher's mind.

On the other hand, Cuebee's process of formulating queries allows a logical interpretation of the question. Queries formulated within Cuebee contain not only the concepts (e.g., oligonucleotides and genes) but also make the relationships explicit in the query (e.g., is homologous to). We show the corresponding query in Fig. 1. The query formulation process in Cuebee leads users to find linkages between concepts by suggesting relationships explicitly. The formulated query is more readable and promotes understanding even to users that are new to *T.cruzi* research or with less domain knowledge. This capability of formulating explicitly structured queries is primarily due to the expressiveness of ontology schemas, which promotes defining the associations between concepts.

8         Amir H. Asiaee1, Prashant Doshi1, Todd Minning2, Satya Sahoo3, Priti Parikh3, Amit Sheth3 and Rick L. Tarleton2

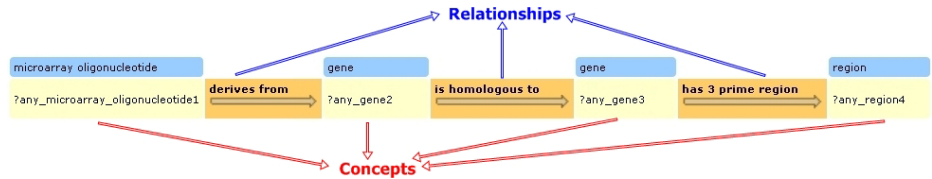

**Fig. 1.** Formulated query for the question, "Which microarray oligonucleotides from homologous genes have 3 prime region primers?" in Cuebee. The concepts and relationships within the query are identified.

### 4.2   Queries at Different Levels of Abstraction

A significant benefit of Cuebee is its ability to allow querying at multiple levels of abstraction. This is beneficial because researchers investigating new hypotheses often ask general questions of their data. In order to illustrate this, consider the following question posed by our life science researchers:

*What genes are used to create any T.cruzi sample?*

Here, *T.cruzi* sample could be of several different types: cloned sample, drug selected sample, transfected sample, and others. Thus, the question is *general* because it targets several different types.

There is no straightforward way to transform this general question into a query using Paige Tools. This is because the relationship between the different types of T.cruzi samples is not explicit in the associated flat database. Currently, researchers translate this question into a query for the strains database that throws up almost all genomic data. Then, three attributes, strain id, strain name and strain status, are analyzed for each data record to ascertain the type of *T.cruzi* sample that the record pertains to. Clearly, this is a tedious approach and relies on much domain knowledge to post process the results. Explicitly linking the different samples would involve redesigning the underlying database requiring multiple additional tables, which leads to reduced efficiency.

On the other hand, Cuebee intuitively encodes the relationships between the different types of samples in the ontology schema: *T.cruzi* sample is a superclass of cloned, drug selected, and transfected samples. A user of Cuebee may translate the question into a triples-based path query as shown in Fig. 2. We note that the query pertains to the class *T.cruzi* sample only (does not include its subclasses in the query). Cuebee's *answer engine* takes advantage of Pellet's inferencing by using SPARQL-DL's extended vocabulary and generates the corresponding query in order to access instances of the class and all its subclasses because subclasses inherit all properties of their superclass. For our example, we see from Fig. 2, that cloned sample -- a subclass of *T.cruzi* sample – appears under the "General Results" tab. There are no results specific to superclass *T.cruzi* sample. Therefore, answering general questions is less dependent on a user's domain expertise in contrast to Paige Tools.

Our observations show that this benefit stands out in two cases: First is when the user is uncertain about which specific concepts relate to her question or she is unaware of more specific concepts. We think that general concepts are easier to identify while formulating the query.

**Fig. 2.** The question "What genes are used to create any T.cruzi sample?" is formulated in Cuebee and cloned sample which is a type of T.cruzi sample appears in the results.

### 4.3 Uniform Query Interface

Ontology-driven approaches such as Cuebee allow a uniform query interface for multiple related datasets; however, Paige Tools offers several interfaces to access the different databases. In order to illustrate this, consider a researcher looking for information on a specific *strain* and a *gene annotation*, which is stored in two different databases, strains and genomic.

Because interfaces in Paige Tools are closely tied to the table schemas of the data that they query, the researcher must load two different interfaces: *strain database* and *gene annotation* query pages. Each form is designed using drop-down lists holding different attribute names from the corresponding table schema and check boxes to give the option of filtering results to the user (see Fig. 3). Notice that the items in the drop-down lists and the check box labels differ across the two interfaces.

(a)

(b)

**Fig. 3.** The gene annotation query and strain database query pages – representing two interfaces of Paige Tools.

Cuebee provides a uniform query interface to the user regardless of which datasets are the target of the questions. Consequently, the process of translating the question into a query does not change with different contexts. Users only need to select a suitable dataset from the drop-down list of datasets. This is enabled by the use of a single, comprehensive ontology schema for all the related datasets. This differs from Paige Tools which, of course, employs different table schemas for each portion of the data. Furthermore, approaches such as Cuebee are usually not tied to a specific ontology but support any ontology designed using the OWL language.

The essential reason behind this flexibility of Cuebee is its use of semantic Web standards that enforce a common data model. Despite different ontologies having distinct concepts and relationships, they all conform to the same data model. Furthermore, standard query languages such as SPARQL are designed to use and be compatible with the data model.

### 4.4  Querying over Multiple Datasets

Often, researchers pose questions that span across different types of data. For example, consider the following question:

*Which genes with log-base-2-ratio greater than 1 have 3 prime region primers*?

In our context, data about genes with log-base-2-ratios is found in the stage transcriptome database while primers with 3 prime regions are found in the strain database. Subsequently, the question spans across two datasets.

In order to translate this question into a query using Paige Tools, researchers utilize two interfaces associated with the different datasets. The question is decomposed into two sequential sub-questions: (a) Which genes have log-base-2-ratio greater than 1; and (b) Which of these genes have 3 prime region primers. Answer to question (a) is found using the *gene annotations query page*. In order to answer (b), a researcher takes the results from (a) and manually looks for the primers in the *gene cloning query page*. While conventional DBMSs do allow queries spanning multiple data sets using joins, Paige Tools does not exploit this partly due to the difficulty of identifying join attributes. Furthermore, facilities to integrate the final results are inadequate.

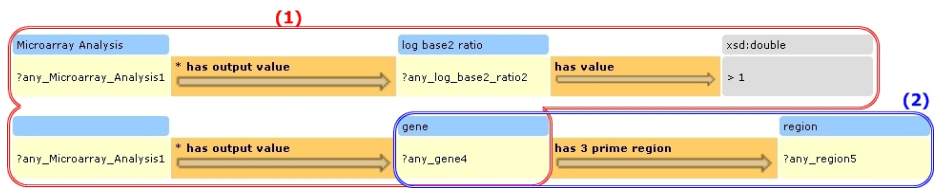

**Fig. 4.** The question, "Which genes with log-base-2-ratio greater than 1 have 3 prime region primers", formulated in Cuebee. Step (1) relates genes and log-base-2-ratio concepts. Step (2) connects genes and 3 prime regions.

On the other hand, Cuebee allows a formulation of the associated query without decomposing it, as illustrated in Fig. 4. A user finds the appropriate concepts and relationships between log-base-2-ratio and gene (Fig. 4 area (1)), and continues to formulate the query by adding the has 3 prime region relationship followed by gene to find the concept region which stores information about region primers (Fig. 4 area (2)). The data related to areas (1) and (2) in Fig. 4, belongs to stage transcriptome and strains datasets, respectively. On formulating the query, Cuebee allows a search over all datasets (as well as individual datasets) – made possible because of a comprehensive ontology for all the data. The solution to the query integrates both datasets thereby facilitating analysis by the researchers with minimal post processing.

## 5     Limitations of Cuebee

We highlight two limitations of knowledge-driven approaches such as Cuebee, which may likely impact its widespread adoption. While ontologies represent a formal model of the domain knowledge, users not well acquainted with the ontology – say, those who have not participated in the engineering of the ontology – feel tied down to its structure. Because the query formulation process is closely linked to the ontology schema, these users express the need to get to know the ontology first. We minimize this through the use of a suggestion engine, which provides suggestions about next possible concepts and associated relationships. Furthermore, our triples-based queries often require users to formulate queries using intermediate concepts and relationships that connect the desired entities in the question. For example, consider the question from Section 3.2: *Which genes are used to create any T.cruzi sample*? As shown in Fig. 2, connecting *gene* and *T.cruzi sample* requires selecting all the intermediate concepts and relationships that link them in the parasite experiment ontology. But users would prefer more abbreviated queries in their daily usage of systems such as Cuebee.

The second limitation is the increased time and space complexity of knowledge-driven systems compared to highly optimized modern DBMSs. This is predominantly due to the ontology inferencing facilities provided by systems such as Pellet, FaCT++ [24], and RacerPro [29]. Furthermore, many of these systems prefer to load the entire ontology and associated instances in main memory. Thus, initial queries consume far more time than later queries. For example, the question shown in Fig. 2 takes about 1.5 minutes to return the results on a high end machine.

## 6     Discussion

We presented a comparative evaluation of two approaches for targeted accessibility to life science data in the context of parasite *T.cruzi* research environment. The first approach, Paige Tools, represents a more conventional approach involving DBMS systems and custom query interfaces. The second approach, Cuebee, belongs to the group of knowledge-driven (ontology-based) query systems.

We described four benefits of Cuebee over Paige Tools. Using Cuebee researchers have the ability to formulate explicitly structured queries which results in a better interpretation of a user's question, compared to Paige Tools. The second benefit of the approach is its capability of generating queries at different levels of abstraction. Using Pellet reasoning, Cuebee takes advantage of the hierarchy of concepts, which is modeled using classes and subclasses in ontologies in order to answer general questions (questions that pertain to high-level concepts with several subtypes). The provision of a uniform query interface is the third benefit of Cuebee over Paige Tools. Using generic data and query models, applicable to any ontology modeled in OWL, enables Cuebee to offer a single query interface for all datasets, compared to Paige Tools which provide different interfaces to access different databases. However, we think that the interfaces in Paige Tools could be made more uniform. The last benefit of approaches such as Cuebee is that they faciliate querying of multiple datasets. In order to answer cross-dataset questions, Cuebee uses Pellet OWL reasoner to integrate multiple datasets with a common ontology schema and expose it as a single query endpoint.

Despite substantial benefits, our approach faces two limitations which may affect widespread usability of Cuebee. We observed that the process of formulating queries relies on a user's knowledge of the structure of ontology schemas. This may be discouraging especially for new users who lack the required knowledge. The second limitation is the computational disadvantage of time and space complexity, similar to many other systems that use ontology inferencing capabilities. Knowledge-driven querying approaches typically consume large amounts of memory and execution time, depending on the size of datasets and the complexity of queries.

We are continuing our comparative observations of both Cuebee and Paige Tools and working toward mitigating the limitations of Cuebee. Specifically, we have formulated a set of 20 questions on which we will objectively evaluate the performance of the two systems. This will provide us with concrete data in order to reinforce our analysis.

We have two avenues to overcome the first limitation mentioned in Section 5: We would like to provide an interface to formulate natural language questions from which to extract the appropriate concepts and relationships. Ramakrishnan [25] and Rosario [26] suggested methods in this regard and to achieve best results we need to tailor these methods to our needs. We could also reduce the query formulation's dependence on the ontology schema by providing better suggestions – early suggestion of concepts that can be reached by another concept in a long path – using the notion of path query discovery [27] [28]. In order to speed up query processing, we plan to save results of previously used queries leading to quicker recall of the results in the future.

**Acknowledgments.**   This research is supported in part by grant number R01HL087795 from the National Heart, Lung and Blood Institute. The content is solely the responsibility of the authors and does not necessarily represent the official views of the National Heart, Lung and Blood Institute or the National Institutes of Health. We thank Pablo Mendes for providing the code for Cuebee and assistance in setting it up, and Alex Tucker, Brandon Ibach and Evren Sirin for providing significant assistance in setting up Pellet and resolving concerns.

14     **Amir H. Asiaee1, Prashant Doshi1, Todd Minning2, Satya** Sahoo3, Priti Parikh3, Amit Sheth3 and Rick L. Tarleton2